\documentclass[conference,10pt]{IEEEtran}
\IEEEoverridecommandlockouts
\usepackage{cite}
\usepackage{amsmath,amssymb,amsfonts}
\usepackage{algorithm}
\usepackage{graphicx}
\usepackage{textcomp}
\usepackage{xcolor}
\usepackage{booktabs}
\usepackage{multirow}
\usepackage{subcaption}    

\usepackage[numbers]{natbib}
\usepackage{hyperref}
\usepackage{pifont}  
\usepackage{threeparttable} 

\usepackage{array}
\newcolumntype{H}{>{\setbox0=\hbox\bgroup}c<{\egroup}@{}}

\usepackage{soul, xcolor}   
\usepackage{tcolorbox}      

\usepackage{float}
\usepackage{subcaption} 

\usepackage[table]{xcolor}

\usepackage{algcompatible}
\algnewcommand\RETURN{\State \textbf{return} }

\definecolor{shapecolor}{rgb}{0.6,0.1,0.0}

\definecolor{commentcolor}{rgb}{0.5,0.5,0.5}

\definecolor{ZZ}{HTML}{0000FF} 
\definecolor{ZG}{HTML}{00AA00} 
\definecolor{SHADE}{HTML}{E6BEF7}

\definecolor{ZZ}{HTML}{000000} 
\definecolor{ZG}{HTML}{000000} 

\definecolor{TITTLE}{HTML}{215fc2}

\usepackage[acronym]{glossaries}
\newacronym{OURMODEL}{MambaNetBurst}{MambaNetBurst}

\begin{document}

\title{
\textcolor{TITTLE}{\acrshort{OURMODEL}: Direct Byte-level Network Traffic Classification \textit{without} Tokenization or Pretraining}
}

\author{
    \IEEEauthorblockN{
    Gayan K. Kulatilleke\IEEEauthorrefmark{1}\IEEEauthorrefmark{2},
    Siamak Layeghy\IEEEauthorrefmark{4},
    Mahsa Baktashmotlagh\IEEEauthorrefmark{3},   
    Marius Portmann\IEEEauthorrefmark{5}
    }
    \IEEEauthorblockA{
        g.kulatilleke@uq.edu.au\IEEEauthorrefmark{1},
        siamak.layeghy@uq.net.au\IEEEauthorrefmark{4},
        m.baktashmotlagh@uq.edu.au \IEEEauthorrefmark{3},        
        marius@itee.uq.edu.au\IEEEauthorrefmark{5}
    }
    \IEEEauthorblockA{
        University of Queensland, Brisbane, Australia
    }
    \IEEEauthorblockA{
        \IEEEauthorrefmark{2}Corresponding author. 
    }    
}

\maketitle

\begin{abstract} 
We present \acrshort{OURMODEL}, a compact tokenizer-free byte-level sequence classifier for network burst classification based on a Mamba-2 backbone. In contrast to most recent strong traffic-classification and intrusion-detection approaches, our method operates directly on raw packet bytes, avoids tokenization, patching, and heavy engineered multimodal representations, and does not require any self-supervised pre-training stage. Given a packet flow, we form a fixed-length burst from the first few packets, embed the resulting byte sequence appending a learnable CLS token, and process it with a stack of residual pre-normalized Mamba-2 blocks for end-to-end supervised classification. Across six public benchmarks spanning encrypted mobile app identification, VPN/Tor traffic classification, malware traffic classification, and IoT attack traffic, \acrshort{OURMODEL} achieves consistently strong results and is competitive with, or outperforms, substantially heavier and often pre-trained baselines. 
Our ablation study shows that preserving byte-level temporal resolution is critical, that early downsampling through striding is consistently harmful, and that moderate state sizes are sufficient for robust generalization. 
We further show that Mamba-2, despite its more constrained transition structure relative to Mamba-1, remains highly effective for packet-byte modeling while providing clear efficiency advantages, particularly in training speed. Overall, our results demonstrate that direct \textit{undiluted}  byte-to-classification learning with compact selective state space models is a 
practical, effective and novel direction for efficient, deployable traffic analysis that bypasses the complexity of pre-training pipelines even over highly optimized linear attention architectures.
\end{abstract}


\begin{table*}[h]
    \centering
    \caption{Comparison of Existing Representation Schemes (updated with pre-training and Mamba details).
        $^{\P}$ PT: Padded/truncated Fixed sizes for headers and payloads.
        $*$ MM: Multimodal input features.}
    \begin{tabular}{lccccccccc}
        \toprule
         Method & Header & Payload & PT\tnote{\P} & MM\tnote{*} & Resolution & Pretraining (training) & Pretrain & Attention & Mamba \\
        \midrule
         PERT~\cite{he2020pert} & \ding{55} & \ding{51} & \ding{55} & \ding{55} & token & MLM (cross-entropy) & \ding{51} & \ding{51} & \ding{55} \\
         ET-BERT~\cite{lin2022bert} & \ding{55} & \ding{51} & \ding{55} & \ding{55} & token & MLM + NSP (cross-entropy) & \ding{51} & \ding{51} & \ding{55} \\
         YaTC~\cite{zhao2023yet} & \ding{51} & \ding{51} & \ding{51} & \ding{55} & patch & MAE reconstruction (MSE) & \ding{51} & \ding{51} & \ding{55} \\
         FlowMAE~\cite{hang2023flow} & \ding{51} & \ding{51} & \ding{55} & \ding{55} & patch & MAE reconstruction (MSE) & \ding{51} & \ding{51} & \ding{55} \\
         NetGPT~\cite{meng2023netgpt} & \ding{51} & \ding{51} & \ding{55} & \ding{55} & token & Next-token (causal LM, cross-entropy) & \ding{51} & \ding{51} & \ding{55} \\
         Lens~\cite{wang2024lens} & \ding{51} & \ding{51} & \ding{55} & \ding{55} & token & Span corruption / seq2seq (cross-entropy) & \ding{51} & \ding{51} & \ding{55} \\
         \midrule
         NetMamba~\cite{wang2024netmamba} & \ding{51} & \ding{51} & \ding{51} & \ding{55} & stride & MAE (cross-entropy) & \ding{51} & \ding{55} & 1 \\
         NetMamba+~\cite{wang2026netmambaplus} & \ding{51} & \ding{51} & \ding{51} & \ding{51} & stride & MAE (Label Distribution-Aware Margin)  & \ding{51} & Partial & 1 \\
         \acrshort{OURMODEL} & \ding{51} & \ding{51} & \ding{51} & \ding{55} & \textcolor{blue}{byte [0\dots255]} & \textcolor{blue}{No-pretraining} (cross-entropy) & \textcolor{blue}{\ding{55}} & \ding{55} & \textcolor{blue}{2} \\
        \bottomrule
    \end{tabular}
    \label{tbl:rep-comp}
\end{table*}

\section{Introduction}
Network traffic classification, which aims to identify the application or service associated with a traffic flow, and network intrusion detection systems (NIDS), which aim to detect malicious or anomalous behavior, are increasingly critical research domains in cybersecurity.
Recent learning-based approaches for traffic analysis have achieved strong results, but many of the most competitive methods rely on large-scale training data, domain-specific pre-training objectives~\cite{lin2022etbert,wang2024netmamba,wang2026netmambaplus}, and sometimes auxiliary metadata~\cite{wang2026netmambaplus}. In the networking and NIDS domains, obtaining such data is costly, time-consuming, and often incomplete or noisy. As a result, many state-of-the-art methods resort to a two-stage strategy in which a model is first pre-trained on traffic-specific self-supervised objectives and then fine-tuned for downstream classification tasks, as in ET-BERT, YaTC, and TrafficFormer~\cite{lin2022etbert,zhao2023yatc,zhou2025trafficformer}.

Despite the inherently sequential nature of network communications, packets and flows, most existing ML pipelines apply some form of early summarization to the raw traffic representation, as seen in Table~\ref{tbl:rep-comp}. 
This is largely motivated by the challenge of processing long input sequences,  large input 'contexts', especially with popular Transformer-based architectures, whose attention mechanism scales quadratically with sequence length, resulting in an exponential increase in computational time. 
Although several efficient attention variants have been proposed, with linear complexity in the attention module~\cite{katharopoulos2020transformers}, at scale the computational cost of a Transformer is dominated by the large feed-forward layers (module) that must also be applied to every input position~\cite{pagnoni2025byte}.
Consequently, tokenization, patching, or striding has generally been treated as necessary for efficient packet sequence classification, as a means to cope with long sequence lengths flow~\cite{pagnoni2025byte}, and to amortize the cost of operating directly on native (byte) resolution~\cite{xue2022byt5}.

However, directly modeling raw bytes offers several advantages. Similar to byte-level modeling in natural language processing, byte-level packet modeling avoids fixed vocabularies, supports arbitrary formats, reduces or eliminates preprocessing complexity, and preserves fine-grained structural information that may otherwise be lost during early aggregation~\cite{xue2022byt5}. 
This is particularly appealing in network traffic analysis, where discriminative cues may occur at very small scales, including protocol-specific fields, payload signatures, and short local patterns.

Because byte-level inputs produce long one-dimensional sequences, architectures with linear-time sequence modeling are especially attractive. 
State space models (SSMs), and Mamba in particular, have emerged as efficient alternatives to Transformers for long-sequence learning, combining linear-time complexity with strong long-range modeling capability~\cite{gu2023mamba}. 
Mamba-2 further improves hardware efficiency by reformulating structured state-space computation into GEMM-friendly kernels via structured state space duality (SSD), resulting in faster training and inference in practice~\cite{dao2024Mamba-2}. 
In parallel, tokenizer-free byte modeling has also shown promise outside networking; ByT5~\cite{xue2022byt5} demonstrated that byte-level transformer models are competitive with their token-level counterparts but significantly more robust to noise, and MambaByte demonstrated that selective SSMs can model raw byte sequences competitively without subword tokenization~\cite{wang2024mambabyte}.

Applying byte-resolution SSMs to network traffic classification is not a direct transfer from language or vision. 
Critical implementation specific decisions must be made regarding ingesting, sequence modeling, and obtaining the classification outputs from the recurrent state or final representations. 
Network packet bytes form long sequential signals with multi-scale structure, including intra-packet byte patterns, packet boundaries, and higher-level exchanges across flows. 
Although many existing pipelines mitigate excessive sequence length through early aggregation, using tokens, patches, or striding, (due to performance and rerource challenges), this may dilute fine-grained discriminative byte-level cues essential for accurate threat detection, malware detection, or intrusion analysis. 
In addition, while Mamba-2 offers clear GPU-efficiency benefits, it imposes structural constraints in its $A$ matrix, raising the question of whether these constraints are expressive enough for network byte sequences. 
To the best of our knowledge, no prior work explores Mamba and Mamba-2 in the context of direct byte-based packet/flow classification in the networking and NIDS domains.

In this work, we ask whether direct modeling of packet bytes can eliminate the need for expensive pre-training, heavy engineered representations, or early downsampling schemes. We introduce \acrshort{OURMODEL}, a compact burst-level classifier that operates directly on raw packet bytes using a Mamba-2 backbone and a lightweight classification head. Our results show that direct byte-to-classification learning is both feasible and highly effective across a range of encrypted traffic, VPN/Tor, IoT, and malware classification benchmarks. Essentially:

\begin{itemize}
    \item We propose \acrshort{OURMODEL}, the first tokenizer-free, pre-training-free, pure byte-level network packet classifier using a compact Mamba-2 backbone.
    \item We demonstrate that Mamba-2's constrained transition matrix (scalar $\times$ identity) is not only sufficient but acts as a beneficial regularizer for packet-byte sequences, while delivering clear GPU efficiency gains.
    \item We show that discriminative traffic representations can be learned directly from packet bytes in a fully supervised setting, eliminating the need for costly pre-training pipelines dominant in the literature.
    \item We evaluate on six public benchmarks (encrypted mobile apps, VPN/Tor, malware, IoT attacks) where \acrshort{OURMODEL} is competitive with or superior to heavier pre-trained baselines.
    \item We provide extensive ablations over embedding design, positional encoding, striding, depth, state size, and Mamba-1 vs.\ Mamba-2, together with detailed forward/backward/eval timing and memory profiles. We release code and models for reproducibility.
\end{itemize}

\section{Related Works}

Network traffic classification, especially for encrypted traffic analysis and intrusion detection, has evolved from handcrafted statistical features to deep ML. Early approaches applied CNNs/GNNs/RNNs to packet payloads or flow sequences~\cite{lo2023xg,sarhan2023doc}. 
Recent work has adopted pre-trained Transformers for packet-byte and flow-level modeling. Methods such as ET-BERT~\cite{lin2022bert} and TrafficFormer learn general traffic representations through self-supervised objectives before being fine-tuned on downstream benchmarks, achieving strong performance across datasets such as USTC-TFC2016~\cite{wang2017ustc} and CICIoT2022~\cite{dadkhah2022ciciot}.
Large Transformer architecture such as GPT-4o and LLaVA have also been using in zero-shot NIDS~\cite{luay2025multimodal}.

A central design choice in traffic representation is the degree of preprocessing applied before the sequence model~\cite{manocchio2024flowtransformer}. 
Pre-trained embeddings have demonstrated strong downstream performance~\cite{zhao2023yet,lin2022bert,zhou2025trafficformer,ptu2024}, but these methods typically depend on carefully engineered input pipelines and pre-training objectives.

Byte-level modeling, mapping from raw data to predictions without any intermediate tokenization or early aggregation, offers a compelling alternative~\cite{wang2024mambabyte}.
By operating on a fixed 256-symbol vocabulary, byte-level modeling avoids subword tokenizer biases. It supports arbitrary data modalities (text, binaries, packets) naturally and directly. 
ByT5~\cite{xue2022byt5} and MambaByte~\cite{wang2024mambabyte} demonstrate the viability of 256-vocabulary modeling on raw bytes, reporting competitive bits-per-byte perplexity and robustness in long-context language modeling.
Raw byte encodings preserve packet structure more directly by encoding headers~\cite{end2end,tfegnn2023zhang, ebsnn2022xi,meng2022packet, tran2025quantifying} as well as payload~\cite{huoh2022flow} as high-dimensional sequences.
Byte based packet classification using CNNs, with no traditional preprocessing, achieves competitive results~\cite{end2end}.
These results suggest that direct byte modeling may also be well suited to packet and flow classification, where inputs are naturally binary and fine-grained local structure is often important.

However, raw packet streams also contain substantial syntactic repetition, such as padding, predictable header fields and session-specific artifacts.
Furthermore, variable packet lengths (IP packets can be between 20 to 65,535 bytes) necessitate (further) batch level padding in ML pipelines.
This increases the computation cost and risk of learning incidental correlations. 
To address this, many recent approaches rely on tokens, patches, strides, or multimodal representations that combine payload bytes~\cite{lin2022etbert,wang2024netmamba,wang2026netmambaplus} and/or packet-level statistics or metadata~\cite{wang2026netmambaplus}.

For long-sequence modeling, state space models (SSMs) have emerged as efficient alternatives to attention-based architectures. Mamba~\cite{gu2023mamba} introduced selective, input-dependent SSM dynamics together with an efficient scan implementation, enabling linear-time sequence processing with strong empirical performance. 
Mamba-2~\cite{dao2024Mamba-2} refined this design through structured state space duality (SSD), reformulating the core computation into matrix-multiplication-friendly operations that improve hardware efficiency while preserving selective sequence modeling behavior.
Relative to Mamba, Mamba‑2 retains Mamba’s selective, input-conditioned SSM dynamics, interpretable as a learned temporal filtering mechanism. 
Its SSD-based reformulation, mapping structured SSMs to attention-like computations, replaces scan-learning on Mamba-1 with matmul-leaning.
Block matmuls and chunked operations are more GMMM-like(General Matrix Multiply) leading to  substantially more efficient core utilization and training throughput, ideal for packet-level byte sequence modeling.
The hardware-optimization comes at a price: Mamba-2 is more constrained/restricted in its core matrix structure, using a 'scalar-times-identity' as its $A$-matrix, whereas Mamba-1 uses a more flexible diagonal structure.

Mamba-based traffic classifiers have only recently begun to appear. NetMamba~\cite{wang2024netmamba}, the pioneering work, replaced Transformer components with a unidirectional Mamba backbone while retaining a carefully designed traffic representation, striding, and a self-supervised pre-training stage. NetMamba+~\cite{wang2026netmambaplus}, by the same team, further extended this line with multi-modal representations (adding inter-arrival times), FlashAttention, and a label-distribution-aware loss. 
ET-Mamba~\cite{xu2025etmamba} proposed a lightweight Mamba model with pre-training and task-specific fine-tuning for encrypted traffic, emphasizing ultra-low parameter count and competitive accuracy on non-VPN datasets.
Further variants include graph-augmented (IDS-GraphMamba~\cite{2025graphmamba}), frequency-aware (HFE-Traffic), and hybrid approaches.

However, existing Mamba-based traffic models rely on engineered inputs such as flow statistics, multimodal features, graph edges, or packet-level aggregates, rather than directly ingesting raw packet bytes.
Supervised pre-training, via contrastive learning~\cite{kulatilleke2022scgc, kulatilleke2023efficient} or masked reconstruction objectives~\cite{wang2024netmamba,zhao2023yatc}, are dominant in traffic classification pipelines.
At the same time, prior work on byte modeling has argued that fixed monotonic patching or striding can be counterproductive, since it ignores disparities in local information density~\cite{pagnoni2025byte} and may break meaningful structures across patch/stride boundaries and/or group (informative) hard bytes together.
Furthermore, despite the possible efficiency gains, no prior work has explored the more constrained Mamba-2 architecture directly on byte-level inputs. 

Motivated by these observations, we study whether the more constrained but efficient Mamba-2 can be applied directly to byte-level packet bursts for network traffic classification, taking advantage of the linear complexity to use undiluted byte resolution, thus avoiding the need for any pre-training.
%
As shown in Table~\ref{tbl:rep-comp}, to the best of our knowledge, \acrshort{OURMODEL} is the first direct byte-to-classification application of Mamba-2 in the network traffic domain, leveraging the (SSM/SSD) architecture’s native efficiency for long byte sequences while retaining the simplicity of a CLS based classification head.

\begin{figure*}[h]
    \centering 
    \includegraphics[width=1.0\textwidth]{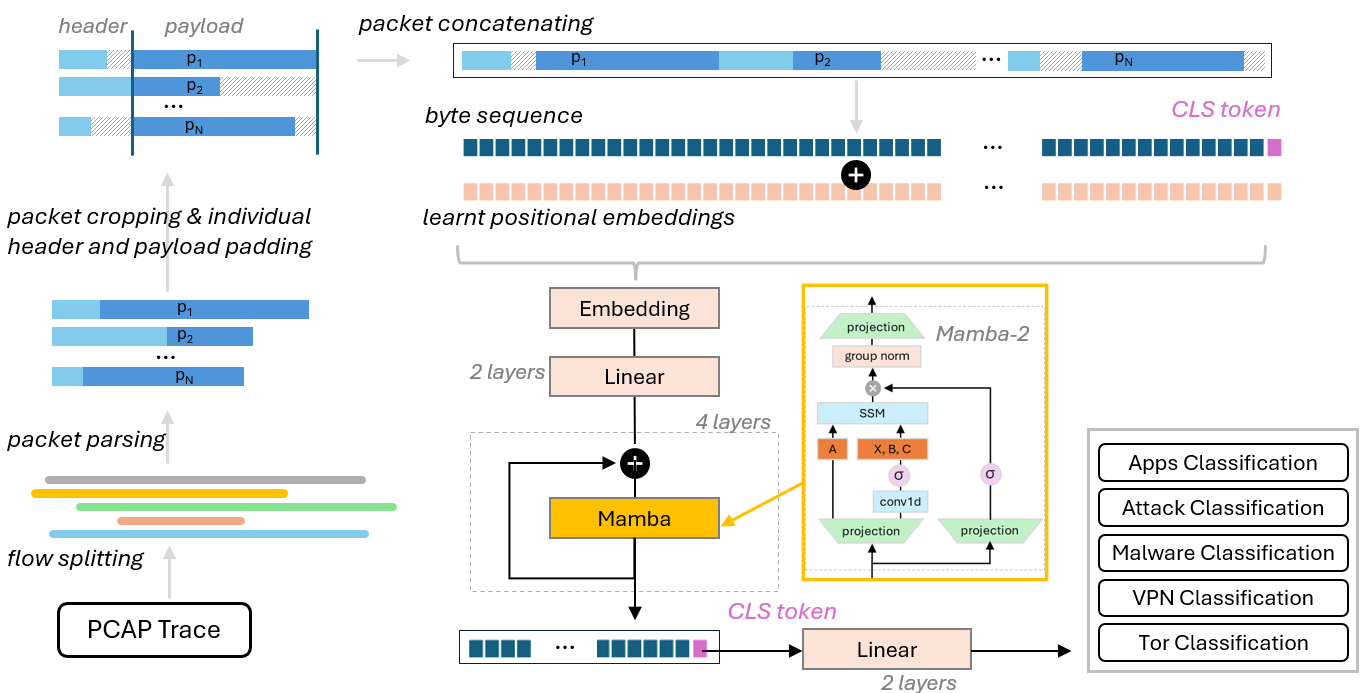}
    \caption{Overview of \acrshort{OURMODEL}. \textbf{Left:} PCAP files are split into 5-tuple flows. The first 5 packets are selected, and each packet is padded or truncated to 320 bytes, yielding a 1600-byte sequence. \textbf{Center:} the sequence is embedded, augmented with positional information and a CLS token, and processed by stacked Mamba-2 layers. The final CLS representation is used for classification. \textbf{Right:} Mamba-2 block architecture~\cite{dao2024Mamba-2}.}
    \label{fig_model}
\end{figure*}

\section{Architecture}

This section describes \acrshort{OURMODEL} (Figure~\ref{fig_model}). The model constructs a fixed-length byte sequence from the first packets of a network flow and feeds this sequence directly into a compact Mamba backbone for supervised classification. Unlike most recent state-of-the-art traffic classifiers, \acrshort{OURMODEL} does not use any self-supervised pre-training stage.

\subsection{Byte sequence construction}
We consider burst-level flow classification. Let the set of all packets be denoted by
\[
P = \{p^1, \dots, p^{|P|}\},
\]
where each packet $p^i = (x^i, y^i, t^i)$ for $i \in [1, |P|]$. Here, $x^i$ denotes the 5-tuple consisting of source IP, destination IP, source port, destination port, and protocol; $y^i$ denotes the packet size in bytes, with $y^i \in (0,\infty)$; and $t^i$ denotes the transmission timestamp in seconds, with $t^i \in [0,\infty)$~\cite{luay2026time}.

Let $F$ denote the set of flows, where each flow $f^k$ is the collection of packets sharing the same 5-tuple. Formally, a unidirectional flow is defined as~\cite{wickramasinghe2025sok}
\begin{equation}
    f^k = \{ p^i \in P \mid x^i = x^k \}.
\end{equation}

Given a flow $f^k$, our goal is to construct a fixed-length burst representation and predict a class label $c \in \{1,\dots,C\}$, corresponding for example to an application, device type, VPN/Tor category, or malware family. We use the first $n$ packets from the flow and retain the first $m$ bytes from each packet, producing a total sequence length of $m \times n$ bytes. This design preserves packet boundaries while focusing on the earliest portion of each packet, which often contains the most informative header and initial payload content~\cite{wickramasinghe2025sok}. In all experiments, we use $n=5$ and $m=320$, yielding a 1600-byte sequence. This is consistent with commonly used truncation settings in prior work~\cite{lin2022etbert}.

\paragraph{Bias control and masking.}
Some datasets contain fields that may introduce label leakage, such as IP addresses, ports, or MAC addresses~\cite{wickramasinghe2025sok}. Following common practice~\cite{lin2022etbert,zhao2023yatc,wang2024netmamba,wang2026netmambaplus}, we mask IP addresses by replacing them with \texttt{0.0.0.0}, remove Ethernet headers during sequence construction, and preserve flow-level train/validation/test splits to avoid overlap across partitions. We also exclude non-IP protocols such as ARP and DHCP. The remaining bytes consist of padded or truncated IP headers together with upper-layer content and payload bytes, matching the benchmark preprocessing conventions.

\subsection{Tokenizer-free byte embedding and projection}
Our input consists of raw network bytes in the range $0,\dots,255$. Each byte value is mapped to a learnable embedding vector of dimension $d_{\text{model}}$:
\begin{equation}
    \mathbf{e}_i = \mathrm{Embed}(x_i) \in \mathbb{R}^{d_{\text{model}}}.
\end{equation}
By default, we use $d_{\text{model}}=256$. This byte-level representation avoids tokenization and allows the model to operate directly on arbitrary packet content.

\subsection{Embedding projection}
A raw embedding lookup provides a learned vector for each byte, but does not itself introduce nonlinear interaction within the feature dimension. To provide a slightly richer per-byte representation before sequence modeling, we apply a lightweight two-layer projection MLP:
\begin{equation}
    \tilde{\mathbf{e}}_i = W_2 \, \phi(W_1 \mathbf{e}_i),
\end{equation}
where $\phi(\cdot)$ is the GELU activation, and $W_1$ and $W_2$ map
$d_{\text{model}} \rightarrow d_{\text{mlp}} \rightarrow d_{\text{model}}$.
This projection is motivated by prior byte-level modeling work such as MambaByte~\cite{wang2024mambabyte}, and provides a simple local feature lifting stage before the Mamba backbone. In practice, we find that this component modestly improves robustness, although the model remains effective without it.

\subsection{Positional encoding (Std pos) and CLS token}
We append a learnable \texttt{CLS} token to the end of the projected byte sequence and add learnable positional embeddings:
\begin{equation}
\mathbf{h}^{(0)} = [\tilde{\mathbf{e}}_1; \dots; \tilde{\mathbf{e}}_T; \mathbf{e}_{\text{CLS}}] + [\mathbf{p}_1; \dots; \mathbf{p}_T; \mathbf{p}_{T+1}],
\end{equation}
where $\mathbf{e}_{\text{CLS}} \in \mathbb{R}^{d_{\text{model}}}$ is the learnable classification token embedding and $\mathbf{p}_i$ denotes the positional embedding for position $i$. The final representation used for classification is the output state at the CLS position, i.e., $\mathbf{h}^{(L)}_{T+1}$.

\subsection{Mamba backbone and classifier head}
We stack $N$ Mamba-2 blocks (default $N=4$) and use the final CLS representation for classification:
\begin{align}
    \mathbf{h}^{(N)} &= \mathrm{MambaStack}(\mathbf{h}^{(0)}) \\
    \hat{\mathbf{y}} &= \mathrm{softmax}(W \mathbf{h}^{(N)}_{\texttt{CLS}})
\end{align}

where $\hat{\mathbf{y}} \in \mathbb{R}^{C}$ denotes the predicted class distribution and $C$ is the number of classes. Training minimizes the cross-entropy loss between $\hat{\mathbf{y}}$ and the ground-truth label $\mathbf{y}$:
\begin{equation}
\mathcal{L}_{\text{cls}} = \mathrm{CrossEntropy}(\hat{\mathbf{y}}, \mathbf{y}).
\end{equation}

The \acrshort{OURMODEL} Mamba-2 block forward pass is outlined in Algorithm 1.

\begin{algorithm*}
\caption{Mamba-2 Block Forward Pass}
\label{alg:mamba2_corrected}
\begin{algorithmic}[1]
\REQUIRE $\mathbf{X} \in \mathbb{R}^{B \times L \times D}$ \COMMENT{Input sequence}
\REQUIRE $\mathbf{A}$ \COMMENT{Learnable static transition parameter}
\ENSURE $\mathbf{Y} \in \mathbb{R}^{B \times L \times D}$ \COMMENT{Output sequence}
\STATE $\mathbf{X}_{norm} \leftarrow \text{LayerNorm}(\mathbf{X})$ \COMMENT{Shape: $(B, L, D)$}
\STATE $\mathbf{z}, \mathbf{x}, \mathbf{B}, \mathbf{C}, \Delta \leftarrow \text{Linear}_{parallel}(\mathbf{X}_{norm})$ \COMMENT{Parallel projections from normalized input}
\STATE $\mathbf{x}_{conv} \leftarrow \text{SiLU}(\text{Conv1d}(\mathbf{x}))$ \COMMENT{Local mixing applied only to main sequence $\mathbf{x}$}
\STATE $\bar{\mathbf{A}}, \bar{\mathbf{B}} \leftarrow \text{Discretize}(\Delta, \mathbf{A}, \mathbf{B})$ \COMMENT{ex: $\bar{\mathbf{A}} = \exp(\Delta \mathbf{A})$, $\bar{\mathbf{B}} = \Delta \odot \mathbf{B}$}
\STATE $\mathbf{y}_{heads} \leftarrow \text{SSD}(\bar{\mathbf{A}}, \bar{\mathbf{B}}, \mathbf{C}, \mathbf{x}_{conv})$ \COMMENT{SSD with $Q=\mathbf{C}, K=\bar{\mathbf{B}}, V=\mathbf{x}_{conv}$}
\STATE $\mathbf{y} \leftarrow \text{reshape}(\mathbf{y}_{heads})$ \COMMENT{Concatenate heads back to $(B, L, E)$}
\STATE $\mathbf{y}_{gated} \leftarrow \mathbf{y} \odot \text{SiLU}(\mathbf{z})$ \COMMENT{Multiplicative gating}
\STATE $\mathbf{Y} \leftarrow \text{Linear}_{out}(\mathbf{y}_{gated}) + \mathbf{X}$ \COMMENT{Output projection and skip connection}
\RETURN $\mathbf{Y}$
\end{algorithmic}
\end{algorithm*}

\section{Experiments}

\subsection{Datasets}
We evaluate on six public benchmarks spanning application identification, IoT device and attack classification, VPN/Tor traffic classification, and malware traffic classification. All splits are performed at the flow level to avoid leakage, and flows from the same capture session or device do not overlap across train, validation, and test partitions when evaluating application identification datasets. For comparison, we use the same splits from prior work~\cite{lin2022etbert,zhao2023yatc,wang2024netmamba}. 
Header bytes retain the IP header with masked IP addresses, and each packet is padded or truncated before five packets are concatenated into a fixed-length flow burst. The evaluated datasets are:
\begin{itemize}
  \item \textbf{CrossPlatform (Android/iOS)} - Encrypted mobile app traffic identification, consisting of  254 and 253 applications respectively~\cite{lin2022etbert}.
  \item \textbf{ISCXVPN2016} - VPN traffic data from 7 communication categories~\cite{drapergil2016vpn}.
  \item \textbf{ISCXTor2016} - Tor traffic data from 8 communication categories~\cite{lashkari2017tor}.
  \item \textbf{USTC-TFC2016} - Malware traffic Classification, distinguishing between malware and benign traffic, of 10 classes each~\cite{wang2017malware}.
  \item \textbf{CICIoT2022} - Attack traffic Classification, such as Denial of Service (DoS)
attacks and brute force attacks.~\cite{dadkhah2022ciciot}.
\end{itemize}


We compare against classical feature-based approaches (AppScanner~\cite{taylor2017robust}, FlowPrint~\cite{van2020flowprint}), supervised deep learning baselines (FS-Net~\cite{liu2019fs}, TFE-GNN~\cite{zhang2023tfe}), pre-trained Transformers (ET-BERT~\cite{lin2022bert}, YaTC~\cite{zhao2023yet}), and a Mamba-based baseline (NetMamba~\cite{wang2024netmamba}). 
YaTC(OF) replaces packet-level and flow-level attention with a global attention module. 
Where indicated, baseline numbers are taken directly from the cited papers.

\subsection{Implementation details and Hyper-parameters}

Unless otherwise stated, we train all models for $120$ epochs using the AdamW optimizer with an initial learning rate of $10^{-3}$ and weight decay of $0.05$. A linear warm-up is applied for the first $10$ epochs, followed by cosine annealing with a minimum learning rate of $10^{-6}$. Training uses mixed-precision with automatic casting and gradient scaling. The default model configuration sets the embedding dimension to $d_{\text{model}}=256$, the encoder depth to $4$ layers, the Mamba state size to $d_{\text{state}}=16$, the classifier hidden dimension to $512$, and dropout to $0.1$. 

For the backbone, we consider four alternatives: \texttt{Mamba-1}, \texttt{Mamba-2}, a standard Transformer encoder, and a linear Transformer encoder. 
We use the official Mamba implementation\footnote{https://github.com/state-spaces/mamba}~\cite{gu2023mamba,dao2024Mamba-2}.
We implement the code for vanilla Transformer~\cite{vaswani2017attention} blocks with quadratic complexity and Linear Transformer~\cite{katharopoulos2020transformers} blocks with linear complexity from respective published works.
For Mamba-based encoders, each residual pre-normalized block uses state-space expansion $d_{\text{state}}=16$, expansion factor $2$, convolution width $d_{\text{conv}}=4$, \texttt{bias=False}, and \texttt{conv\_bias=True}; for Mamba-2, we additionally set the head dimension to $64$, whereas Mamba-1 does not use a head-dimension parameter.
For the Transformer-based variants, we use $4$ attention heads; the Transformer and linear Transformer feed-forward dimension is set to $256$ in the encoder blocks. 

The sequence length is $1600$ bytes.
A learnable \texttt{[CLS]} token is appended to the end of the sequence, and learnable positional embeddings are added to all tokens. 
We trained on a Nvidia RTX 3090, 23.54 GiB with a batch size of $128$ for Mamba-based models, and a much smaller batch size of $32$ for transformer variants due to their higher memory cost. 
We do not use any pre-training. 
During evaluation, we report accuracy (AC), precision (PR), recall(RC), and macro-F1(F1).

\section{Results}

\subsection{Network packet classification}
Tables~\ref{tb:overall-1}--\ref{tb:overall-2} summarize results across six publicly available datasets. Baseline entries are reported from NetMamba~\cite{wang2024netmamba} where indicated. 
Across the six evaluated datasets, \acrshort{OURMODEL} achieves high macro-F1 in the main results tables, including very strong performance on ISCXTor2016 and USTC-TFC2016 (Table~\ref{tb:overall-2}) and strong results on CrossPlatform (Android/iOS) (Table~\ref{tb:overall-1}). Importantly, these results are obtained without any pretraining stage.

\begin{table*}[h]
  \centering
  \caption{Comparison Results on CrossPlatform(Android), CrossPlatform(iOS) and CICIoT2022. 
  Source baselines:~\cite{wang2024netmamba}.}
  \resizebox{\textwidth}{!}{
  \begin{tabular}{l|rr|cccc|cccc|cccc}
    \toprule
    \multirow{3}{*}{Method} 
    & \multicolumn{2}{c|}{} & \multicolumn{8}{c|}{Encrypted mobile app traffic Classification}  & \multicolumn{4}{c}{Attack traffic Classification} \\
    
    & \multicolumn{2}{c|}{Params(M)} 
    & \multicolumn{4}{c|}{CrossPlatform(Android)\cite{ren2019international}} & \multicolumn{4}{c|}{CrossPlatform(iOS)\cite{ren2019international}} & \multicolumn{4}{c}{CICIoT2022 \cite{dadkhah2022towards}} \\    
    
    \cline{2-15} \noalign{\vskip 0.5mm}
     & PT & FT & AC & PR & RC & F1 & AC & PR & RC & F1 & AC & PR & RC & F1 \\
     \midrule
     AppScanner~\cite{taylor2017robust} & - & - & 0.1626 & 0.1646 & 0.1456 & 0.1413 & 0.1718 & 0.1400 & 0.1440 & 0.1283 & 0.7556 & 0.8093 & 0.7244 & 0.6938 \\
     FlowPrint~\cite{van2020flowprint} & - & - & 0.8739 & 0.8941 & 0.8739 & 0.8700 & 0.8712 & 0.8687 & 0.8712 & 0.8603 & 0.5820 & 0.4164 & 0.5820 & 0.4643 \\
     \midrule
     FS-Net~\cite{liu2019fs} & - & 5.3 & 0.0147 & 0.0023 & 0.0147 & 0.0034 & 0.0293 & 0.0014 & 0.0293 & 0.0025 & 0.5747 & 0.3800 & 0.5747 & 0.4216 \\
     TFE-GNN~\cite{zhang2023tfe} & - & 44.3 & 0.8141 & 0.8308 & 0.8141 & 0.8067 & 0.8241 & 0.8326 & 0.8241 & 0.8130 & \cellcolor{lightgray}1.000 & \cellcolor{lightgray}1.000 & \cellcolor{lightgray}1.000 & \cellcolor{lightgray}1.000 \\
     \midrule
     ET-BERT~\cite{lin2022bert} & 187.4 & 136.4 & 0.8743 & 0.8913 & 0.8743 & 0.8786 & 0.9105 & 0.8809 & 0.9105 & 0.8850 & 0.9937 & 0.9938 & 0.9937 & 0.9937 \\
     YaTC(OF)~\cite{zhao2023yet} & 2.3 & 2.1 & 0.9076 & 0.9107 & 0.9076 & 0.9077 & 0.9263 & 0.9282 & 0.9263 & 0.9264 & 0.9949 & 0.9949 & 0.9949 & 0.9949 \\
     YaTC~\cite{zhao2023yet} & 2.3 & 2.1 & 0.8952 & 0.8989 & 0.8952 & 0.8952 & 0.9270 & 0.9296 & 0.9270 & 0.9272 & 0.9974 & 0.9975 & 0.9974 & 0.9974 \\
     \midrule
     NetMamba~\cite{wang2024netmamba} & 2.2 & 1.9 & 0.9094 & 0.9133 & 0.9094 & 0.9096 & 0.9301 & 0.9327 & 0.9301 & 0.9305 & 0.9928 & 0.9931 & 0.9928 & 0.9929 \\
    \midrule
    \acrshort{OURMODEL} & NA & 2.7/\textbf{2.5} 
    & \cellcolor{lightgray}0.9860 & \cellcolor{lightgray}0.9838 & \cellcolor{lightgray}0.9831 & \cellcolor{lightgray}0.9824
    & \cellcolor{lightgray}0.9900 & \cellcolor{lightgray}0.9837 & \cellcolor{lightgray}0.9875 & \cellcolor{lightgray}0.9851 
    & 0.9974 & 0.9967 & 0.9964 & 0.9966 \\
    \bottomrule
    
  \end{tabular}
  }
  \label{tb:overall-1}
\end{table*}

\begin{table*}[h]
  \centering
  \caption{Comparison Results on ISCXTor2016, ISCXVPN2016 and USTC-TFC2016. 
  Source baselines:~\cite{wang2024netmamba}.}
  \resizebox{\textwidth}{!}{
  \begin{tabular}{l|rr|cccc|cccc|cccc}
    \toprule
    \multirow{3}{*}{Method} 

    & \multicolumn{2}{c|}{} & \multicolumn{4}{c|}{Tor traffic Classification} & \multicolumn{4}{c|}{VPN traffic Classification} & \multicolumn{4}{c}{Malware traffic Classification} \\
    
    & \multicolumn{2}{c|}{Params(M)} & \multicolumn{4}{c|}{ISCXTor2016 \cite{lashkari2017characterization}} & \multicolumn{4}{c|}{ISCXVPN2016 \cite{gil2016characterization}} & \multicolumn{4}{c}{USTC-TFC2016 \cite{wang2017malware}} \\
    
    \cline{2-15} \noalign{\vskip 0.5mm}
     & PT & FT & AC & PR & RC & F1 & AC & PR & RC & F1 & AC & PR & RC & F1 \\
     \midrule
     AppScanner~\cite{taylor2017robust} & - & - & 0.4034 & 0.2850 & 0.2149 & 0.2113 & 0.7643 & 0.8047 & 0.7045 & 0.7256 & 0.6998 & 0.8591 & 0.6062 & 0.6633 \\
     FlowPrint~\cite{van2020flowprint} & - & - & 0.1316 & 0.0173 & 0.1316 & 0.0306 & 0.9666 & 0.9733 & 0.9666 & 0.9681 & 0.7992 & 0.7745 & 0.7992 & 0.7755 \\
     \midrule
     FS-Net~\cite{liu2019fs} & - & 5.3 & 0.7020 & 0.7010 & 0.7020 & 0.6999 & 0.7023 & 0.7487 & 0.7023 & 0.6660 & 0.4381 & 0.2011 & 0.4381 & 0.2672 \\
     TFE-GNN~\cite{zhang2023tfe} & - & 44.3 & 0.7692 & 0.8030 & 0.7692 & 0.7618 & 0.8428 & 0.8508 & 0.8428 & 0.8447 & 0.9747 & 0.9747 & 0.9747 & 0.9734 \\
     \midrule
     ET-BERT~\cite{lin2022bert} & 187.4 & 136.4 & 0.9967 & 0.9967 & 0.9967 & 0.9967 & 0.9566 & 0.9566 & 0.9566 & 0.9565 & 0.9910 & 0.9911 & 0.9910 & 0.9910 \\
     YaTC(OF)~\cite{zhao2023yet} & 2.3 & 2.1 & 0.9986 & 0.9986 & 0.9986 & 0.9986 & 0.9805 & 0.9808 & 0.9805 & 0.9806 & 0.9960 & 0.9955 & 0.9960 & 0.9957 \\
     YaTC~\cite{zhao2023yet} & 2.3 & 2.1 & 0.9959 & 0.9959 & 0.9959 & 0.9959 & \cellcolor{lightgray}0.9848 & 0.9849 & 0.9848 &  0.9848 & 0.9972 & \cellcolor{lightgray}0.9976 & \cellcolor{lightgray}0.9972 & \cellcolor{lightgray}0.9970 \\
    \midrule
    NetMamba~\cite{wang2024netmamba} & 2.2 & 1.9 & 0.9986 & 0.9986 & 0.9986 & 0.9986 & 0.9805 & 0.9808 & 0.9805 & 0.9806 & 0.9960 & 0.9957 & 0.9960 & 0.9957 \\
    \midrule

    \acrshort{OURMODEL} & NA & 2.7/\textbf{2.5} 
    & \cellcolor{lightgray}0.9993 & \cellcolor{lightgray}0.9991 & \cellcolor{lightgray}0.9990 & \cellcolor{lightgray}0.9990
    & 0.9834 & \cellcolor{lightgray}0.9884 & \cellcolor{lightgray}0.9859 & \cellcolor{lightgray}0.9871
    & \cellcolor{lightgray}0.9995 & 0.9964 & 0.9949 & 0.9954
    \\
    
    \bottomrule
  \end{tabular}
  }
  \label{tb:overall-2}
\end{table*}

\subsection{Ablations}
Table~\ref{tb:ablations-1} summarizes our ablation study across six byte-based packet sequence benchmarks. Unless otherwise stated, all variants use a Mamba-2 backbone with $d_{\text{model}}{=}256$, $d_{\text{state}}{=}16$, $L{=}4$ layers, $d_{\text{mlp}}{=}512$, and $d_{\text{conv}}{=}4$. 
We report per-dataset macro-F1, together with the mean (AVG), worst-case (MIN), best-case (MAX), and cross-dataset variance (VAR). 
Input bytes are represented using one of three embedding strategies: (i) learned byte embeddings with a residual projection (\texttt{byte}), (ii) learned byte embeddings without projection (\texttt{ByteEmbedNoProj}), or (iii) stride-based convolutional embedding (\texttt{stride}) with stride size $4$.
Overall, the ablations indicate that (i) the task is highly solvable with compact selective SSM backbones, (ii) preserving byte-level temporal resolution is critical, and (iii) moderate state capacity and adequate model width yield the most robust generalization across datasets.

\begin{table*}[h]
  \centering
  \caption{Ablations. Unless specified: backbone=Mamba-2, d\_state=16, d\_model=256, layers=4, d\_mlp=512, d\_conv=4. \textbf{A:} indicates Accuracy and \textbf{F1:} is macro F1-score.
  $(64/64/2)$ and $(32/32/2)$ are compact variants where the values indicate d\_model,mlp,layers.
  * Indicates linear transformer using flash attention 2, which only supports fp16/bf16}
  \resizebox{\textwidth}{!}{
  \begin{tabular}{l| cccccc | rrrr }
    \toprule
Method  & ISCXVPN2016 & ISCXTor2016 & USTC-TFC2016 & CICIoT2022 & CP(Android) & CP(iOS) & AVG      & MIN    & MAX    & VAR $\downarrow$ \\
    \midrule

    \rowcolor{blue!8} A:Without pos enc &    0.9870 &   0.9986 &   0.9998 &   0.9964 &   0.9880 &   0.9864 \\
    \rowcolor{blue!8} A:Std pos &   0.9834 &   0.9993 &   0.9995 &   0.9974 &   0.9860 &   0.9900 \\
    \rowcolor{blue!8} A:Std pos (Mamba-1)  &   0.9834 &   0.9966 &   0.9997 &   0.9974 &   0.9800 &   0.9822 \\ 
    
    \rowcolor{green!15} A:Stride (4)  &   0.9704 &   0.9979 &   0.9991 &   0.9969 &   0.9704 &   0.9603\\
    \rowcolor{green!15} A:Stride (4,Mamba-1) &   0.9812 &   0.9966 &   0.9938 &   0.9954 &   0.9729 &   0.9774\\
    \rowcolor{green!15} A:Stride (2) & 0.9812 & 0.9986 & 0.9982 & 0.9944 & 0.9738 & 0.9781 \\
    \rowcolor{green!15} A:Stride (2, Mamba-1) & 0.9718 & 0.9925 & 0.9992 & 0.9836 & 0.9665 & 0.9812 \\
    
    A:No emb proj &   0.9855 &   0.9959 &   0.9997 &   0.9964 &   0.9824 &   0.9869 \\
    \rowcolor{gray!15} A:2 layers &   0.9790 &   0.9973 &   0.9994 &   0.9979 &   0.9820 &   0.9864 \\
    \rowcolor{gray!15} A:1 layers &   0.9827 &   0.9979 &   0.9994 &   0.9959 &   0.9822 &   0.9893\\
    
    \rowcolor{orange!20} A: d\_state 32 & 0.9827& 0.9973& 0.9994& 0.9954& 0.9856& 0.9883\\
    \rowcolor{orange!20} A: d\_state 64 & 0.9812& 0.9973& 0.9994& 0.9964& 0.9869& 0.9883\\
    \rowcolor{orange!20} A: d\_state 128 & 0.9769& 0.9973& 0.9997& 0.9969& 0.9867& 0.9800\\
    A: Compact(\textcolor{blue}{(64/64/2)}) & 0.9725& 0.9959& 0.9997& 0.9959& 0.9747& 0.9812 \\
    A: Compact(32/32/2) &  0.9624&  0.9966&  0.9995&  0.9933&  0.8982&  0.9303 \\
    
    \rowcolor{purple!20} A: Transformer   & 0.9783 & 0.9979 & 0.9994 & 0.9969 & 0.9280 & 0.9857 \\
    A: Linear Tr(flash att 2)* & 0.9913 & 0.9959 & 0.9997 & 0.9974 & 0.9895 & 0.9912 \\
    A: Linear Tr(flash att 2)*(\textcolor{blue}{(64/64/2)}) & 0.9877 & 0.9966 & 0.9995 & 0.9944 & 0.9369 & 0.9719 \\
    
    \midrule
    \rowcolor{blue!8} F1:Without pos enc               & 0.9859      & 0.9980       & 0.9978       & 0.9952     & 0.9847      & 0.9810   & 0.9904 & 0.9810  & 0.9980  & 5.53E-05     \\
    \rowcolor{blue!8} F1:Std   pos                       & 0.9871      & 0.9990*       & 0.9954       & 0.9966     & 0.9824      & 0.9851  & 
    0.9909 & 0.9824 &	0.9990 &	4.77E-05 \\

    \rowcolor{blue!8} F1:Std   pos (Mamba-1)              & 0.9828      & 0.9957      & 0.9956       & 0.9966     & 0.9769      & 0.9770   & 0.9874 & 0.9769 & 0.9966 & 9.21E-05     \\
    
    \rowcolor{green!15} F1:Stride(4)                       & 0.9524      & 0.9972      & 0.9950        & 0.9961     & 0.9662      & 0.9561  & 0.9772 & 0.9524 & 0.9972 & 4.51E-04     \\
    \rowcolor{green!15} F1:Stride(4,Mamba-1)                & 0.9732      & 0.9955      & 0.9614       & 0.992      & 0.9703      & 0.9731  & 0.9776 & 0.9614 & 0.9955 & 1.77E-04     \\
    \rowcolor{green!15} F1:Stride (2) & 0.9727 & 0.9980 & 0.9881 & 0.9902 & 0.9709 & 0.9739 & 0.9823 &	0.9709 & 	0.9980 &	1.27E-04 \\
    \rowcolor{green!15} F1:Stride (2, Mamba-1) & 0.9586 & 0.9914 & 0.9930 & 0.9790 & 0.9645 & 0.9772 &
    0.9773 &	0.9586 &	0.9930 & 1.92E-04 \\

    F1:No   emb proj                   & 0.9838      & 0.9952      & 0.9977       & 0.9955     & 0.9799      & 0.9841  & 0.9894 & 0.9799 & 0.9977 & 5.79E-05     \\
    \rowcolor{gray!15} F1:2   layers                      & 0.9782      & 0.9960       & 0.9944       & 0.9971     & 0.9794      & 0.9816  & 0.9878 & 0.9782 & 0.9971 & 7.97E-05     \\
    \rowcolor{gray!15} F1:1   layers                      & 0.9722      & 0.9970       & 0.9945       & 0.9944     & 0.9796      & 0.9845  & 0.9870 & 0.9722 & 0.9970  & 9.82E-05     \\
    
    \rowcolor{orange!20} F1:   d\_state 32                  & 0.9791      & 0.9960       & 0.9912       & 0.9938     & 0.9837      & 0.9833  & 0.9879  & 0.9791 & 0.9960  & 4.55E-05     \\
    \rowcolor{orange!20} F1:   d\_state 64                  & 0.9769      & 0.9960       & 0.9950        & 0.9952     & 0.9838      & 0.9834  & 0.9883 & 0.9769 & 0.9960  & 6.52E-05     \\
    \rowcolor{orange!20} F1:   d\_state 128                 & 0.9612      & 0.9960       & 0.9977       & 0.9963     & 0.9841      & 0.9743  & 0.9849 & 0.9612 & 0.9977 & 2.18E-04     \\
    
    F1: Compact(\textcolor{blue}{(64/64/2)}) & 0.9618      & 0.9942      & 0.9956       & 0.9946     & 0.9711      & 0.9755  & 0.9821 & 0.9618 & 0.9956 & 2.12E-04     \\
    F1: Compact(32/32/2) & 0.9476      & 0.9950       & 0.9963       & 0.9917     & 0.9060       & 0.9322  & 0.9615 & 0.9060  & 0.9963 & 1.48E-03 \\  
    \rowcolor{purple!20} F1: Transformer & 0.9766 & 0.9972 & 0.9933 & 0.9960 & 0.9289 & 0.9810 & 0.9788 & 0.9289 &	0.9972 & 6.69E-04 \\
    F1: Linear Tr(flash att 2)* & 0.9920 & 0.9945 & 0.9977 & 0.9966 & 0.9871 & 0.9868 & 0.9925 & 0.9868	& 0.9977 & 2.19E-05 \\
    F1: Linear Tr(flash att 2)*\textcolor{blue}{(64/64/2)} & 0.9866 & 0.9954 & 0.9933 & 0.9927 & 0.9367 & 0.9673
    & 0.9787 &	0.9367	& 0.9954 &	5.29E-04 \\
    \bottomrule
  \end{tabular}
  }
  \label{tb:ablations-1}
\end{table*}

\noindent \textbf{Positional embeddings are not dominant.} 
We compare standard positional encoding (\texttt{Std pos}) against removing positional encoding entirely (\texttt{Without pos enc}). While both configurations attain close scores, \texttt{Std pos} achieves better a mean (AVG $0.9909$ vs. $0.9904$), better worst-case performance (MIN $0.9824$ vs. $0.9810$) and lower variance (VAR \ $4.77\times10^{-5}$ vs. $5.53\times10^{-5}$).

This suggests that, for byte-based packet sequences, the ordering bias induced by causal convolution and recurrent state-space dynamics already provides strong sequential structure enabling robust discrimination even without added (absolute) positional features. Positional embeddings remain useful primarily because they improve consistency, as indicated by lower variance, across datasets.
%
Practically, this indicates that packet-byte classification can be dominated by local and local-to-intermediate patterns (meso-scale motifs) \footnote{intermediate structural or functional patterns that exist between the micro- and macro-scales, acting as building blocks for complex system behavior} (e.g., protocol headers, fields, short repeated patterns, delimiters, length fields, checksum starts, common byte sequences in packet structures (amounting to tens to hundreds of bytes), or characteristic 'motifs' that appear in many packets of the same class such as TCP flags + options patterns, HTTP method + version strings.) rather than requiring absolute global positional anchoring.
The lower variance is the key reason to justify the use of positional encodings in \acrshort{OURMODEL}.

\noindent \textbf{Mamba-2 is more consistent than Mamba-1 under matched settings.}
We replace Mamba-2 with Mamba-1 under otherwise matched settings (\texttt{Std pos (Mamba-1)}). Mamba-2 yields higher mean performance (AVG $0.9909$ vs.\ $0.9874$), a better worst-case score (MIN $0.9824$ vs.\ $0.9769$), and substantially lower variance (VAR $4.77\times10^{-5}$ vs.\ $9.21\times10^{-5}$).
These results indicate that the more constrained structured dynamics of Mamba-2 are not only sufficient for byte-level packet classification, but may also act as an implicit regularizer in this setting.
The improved consistency across datasets suggests that the constraints in Mamba-2 may act as an implicit regularizer for this modality, where overly flexible per-channel dynamics (as in Mamba-1) may not be necessary to capture the discriminative structure present in the data.

\noindent \textbf{Early downsampling is harmful.} 
To test sensitivity to temporal resolution, we apply striding with factor 4 (\texttt{Stride(4)}). This change produces the largest degradation among all single-factor ablations, reducing AVG from $0.9909$ to $0.9772$ and substantially increasing variance to $4.51\times10^{-4}$. The worst-case dataset drops to MIN $0.9524$, indicating that some benchmarks critically depend on fine-grained byte order and short-range structure which is lost under downsampling.
%
Using Mamba-1 with the same striding slightly improves robustness relative to strided Mamba-2, indicating that the additional flexibility of Mamba-1 can partially compensate, but performance remains well below the non-strided baseline. This confirms that preserving  fine-grained byte-level temporal fidelity is essential for robust classification across heterogeneous datasets. Consequently, we recommend avoiding striding for accuracy-critical settings and only employing it when compute constraints require downsampling.

\noindent \textbf{Moderate depth is sufficient, but additional layers improve robustness.} 
Reducing depth from $L{=}4$ to $L{=}2$ (\texttt{2 layers}) yields a modest drop in mean performance (AVG $0.9878$), indicating that the task does not require deep models to achieve strong results.
However, the worst-case performance declines as depth decreases, suggesting that additional layers mainly improve robustness rather than average accuracy. 
This is consistent with the view that packet-byte classification contains strong local discriminative cues that shallow sequence models can already capture effectively.
Capacity reductions that jointly shrink width and depth (e.g., \texttt{Compact...}) cause a pronounced collapse, driven by the CrossPlatform datasets.
These results suggest that the modality benefits from additional composition, but does not require substantial depth to attain high F1. Additional layers primarily improve robustness (worst-case performance) rather than dramatically increasing average accuracy, consistent with the view that packet-byte classification contains strong local cues that are extractable with relatively shallow sequence models.

\noindent \textbf{Large state sizes are unnecessary.}
Byte based packet classification does not need a large latent dynamical memory.
Increasing $d_{\text{state}}$ from $16$ to $32$ or $64$ produces only minor changes in mean performance, while further increasing to $128$ degrades both the average and worst-case scores and increases variance. 
This pattern implies that byte-based packet classification does not strongly benefit from very large latent dynamical memory, and that excessive state capacity may be detrimental under fixed training and regularization budgets (e.g., by overfitting dataset-specific long-range artifacts or by shifting capacity away from the short- and mid-range structures that dominate discrimination). For this modality, moderate $d_{\text{state}}$ (16--64) appears to be a favorable operating range.

\noindent \textbf{Adequate model width is important for robust generalization.}
Compact variants that reduce $d_{\text{model}}$, $d_{\text{mlp}}$, and depth simultaneously remain competitive up to a point. The medium-sized variant with $d_{\text{model}}=64$ and two layers preserves strong performance, whereas the smallest variant with $d_{\text{model}}=32$ degrades noticeably, particularly on the CrossPlatform datasets. 
These findings underscore that sufficient representation width is crucial for robust generalization in packet-byte classification, particularly for datasets with higher intra-class variability or weaker signature patterns. While the smallest model remains effective on several benchmarks, it lacks the channel capacity required to consistently encode the diverse discriminative cues present in the CP datasets.

\noindent \textbf{Embedding Projection helps}
Removing the embedding projection (\texttt{No emb proj}) marginally affects performance (AVG $0.9894$ vs.\ $0.9909$ baseline), with comparable worst-case and variance. This indicates that the model can learn effective representations directly from the byte embedding stream in this configuration, while the projection layer mainly offers a small robustness gain.

\noindent \textbf{In summary, }
across all ablations, we identify three consistent trends. First, preserving byte-level temporal resolution is critical: striding causes the largest and most variance-increasing degradation, implying that fine-grained byte patterns carry essential discriminative signal. Second, Mamba-2 provides the best overall robustness in the non-strided setting, improving mean and worst-case F1 while reducing variance across datasets relative to Mamba-1. Third, hyperparameter scaling exhibits modality-specific optima: moderate $d_{\text{state}}$ (16--64) and adequate $d_{\text{model}}$ are more important than increasing state size aggressively. 
Finally, we note that $d_{\text{conv}}$ and expansion factor (\texttt{expand}) are held constant in Table~\ref{tb:ablations-1}; thus, conclusions about these parameters are necessarily indirect and should be validated with targeted sweeps in future work.

\subsection{Mamba-1 vs Mamba-2 scaling} \label{sec:scaling}
Table~\ref{tab:mamba-scaling} reports timing and memory for matched Mamba-1 vs Mamba-2 configurations across batch sizes (RTX 3090). For the 2.5--2.7M parameter setting, Mamba-2 consistently reduces forward and (especially) backward time relative to Mamba-1, while memory usage remains of similar magnitude across batch sizes. The table also shows an out-of-memory condition at batch 256 for Mamba-2 in this specific configuration.

\begin{table*}[h]
    \centering
    \small
    \caption{Scaling Comparison: Mamba-1 vs Mamba-2 at Larger Batch Sizes. (times in ms, peak memory in MiB.) \\
    Averaged over 10 runs, each with a warm-up of 10. Implementation: mamba\_ssm. OOM indicates out of memory on RTX 3090, 23.54GiB.}
    \label{tab:mamba-scaling}
    
    \begin{tabular}{l rrrrrrr}
    \toprule
    \multirow{2}{*}{\textbf{Architecture}} & 
    \multirow{2}{*}{\textbf{Batch}} & \multicolumn{3}{c}{\textbf{Time (ms)}} & \multicolumn{3}{c}{\textbf{Peak Memory (MiB)}} \\
    \cmidrule(lr){3-5} \cmidrule(lr){6-8}
    & & {Forward} & {Back} & {Eval} & {Forward} & {Back} & {Eval} \\
    \midrule
    \multirow{6}{*}{Mamba 1 (2.7M params)}
    & 8      &  6.38 &  18.79 &   6.34 &   620.50 &   754.29 &   246.89 \\
    & 16     & 12.16 &  43.74 &  16.68 &  1197.66 &  1459.44 &   448.36 \\
    & 32     & 23.71 &  98.90 &  29.33 &  2353.65 &  2873.82 &   856.34 \\
    & 64     & 51.35 & 205.56 &  53.52 &  4675.68 &  5714.95 &  1674.65 \\
    & 128    &107.07 & 414.15 & 108.51 &  9324.30 & 11401.78 &  3311.33 \\
    & 256    &258.84 & 824.62 & 415.79 & 18629.28 & 22784.18 &  6586.31 \\
    \midrule
    \multirow{6}{*}{Mamba 2 (2.5M params)}
    & 8      &  5.65 &  12.88 &   5.91 &   618.63 &   798.47 &   254.36 \\
    & 16     & 10.85 &  23.36 &  11.23 &  1175.65 &  1504.09 &   451.32 \\
    & 32     & 21.28 &  44.89 &  20.86 &  2300.96 &  2954.69 &   837.57 \\
    & 64     & 42.87 &  86.60 &  41.84 &  4554.12 &  5860.71 &  1613.82 \\
    & 128    & 86.50 & 166.61 &  85.55 &  9061.93 & 11673.69 &  3167.67 \\
    & 256    & OOM   & OOM    & OOM    & OOM      & OOM      & OOM     \\
    \bottomrule
\end{tabular}
\end{table*}

The backward-pass speedup ($\sim$50\% on average) is especially important in practice since it directly accelerates training/fine-tuning.
Across the evaluated settings, Mamba-2 delivers substantially faster backpropagation.
Forward and evaluation times are also consistently better with Mamba-2 (8–20\% range).
Memory usage is marginally lower for Mamba-2 at the same batch sizes ($\approx 2\%$ savings), despite the OOM at batch 256.
For the backward pass, Mamba-1 uses a recompute approach to lower memory usage at the cost of slower processing, while Mamba-2 stores more intermediates, speeding up backward passes but increasing memory consumption, potentially causing OOM. 
These results are consistent with the design motivation of Mamba-2, which reformulates the structured state-space computation into more hardware-efficient SSD-based operations~\cite{dao2024Mamba-2}. 
Mamba-2’s higher backward memory usage comes from storing chunk intermediates or additional buffers. Its SSD formulation improves computational efficiency but leads to higher memory overhead due to intermediate state storage. While both models have comparable complexity, Mamba-2’s blockwise decomposition and multi-head SSM provide efficiency at larger batches despite memory limits.

In our experiments, this leads to clearly better training throughput without compromising classification performance. Notably, scaling behaves more efficiently for Mamba-2. We provide further batch-wise scaling details in the Appendix Tables~\ref{tab:mamba-scaling-d64-v2} and Table~\ref{tab:mamba-scaling-d256-v2}.



\subsection{Mamba vs others on Accuracy (F1) to Inference time}
\begin{figure*}[h]
    \centering 
    \includegraphics[width=0.98\textwidth]{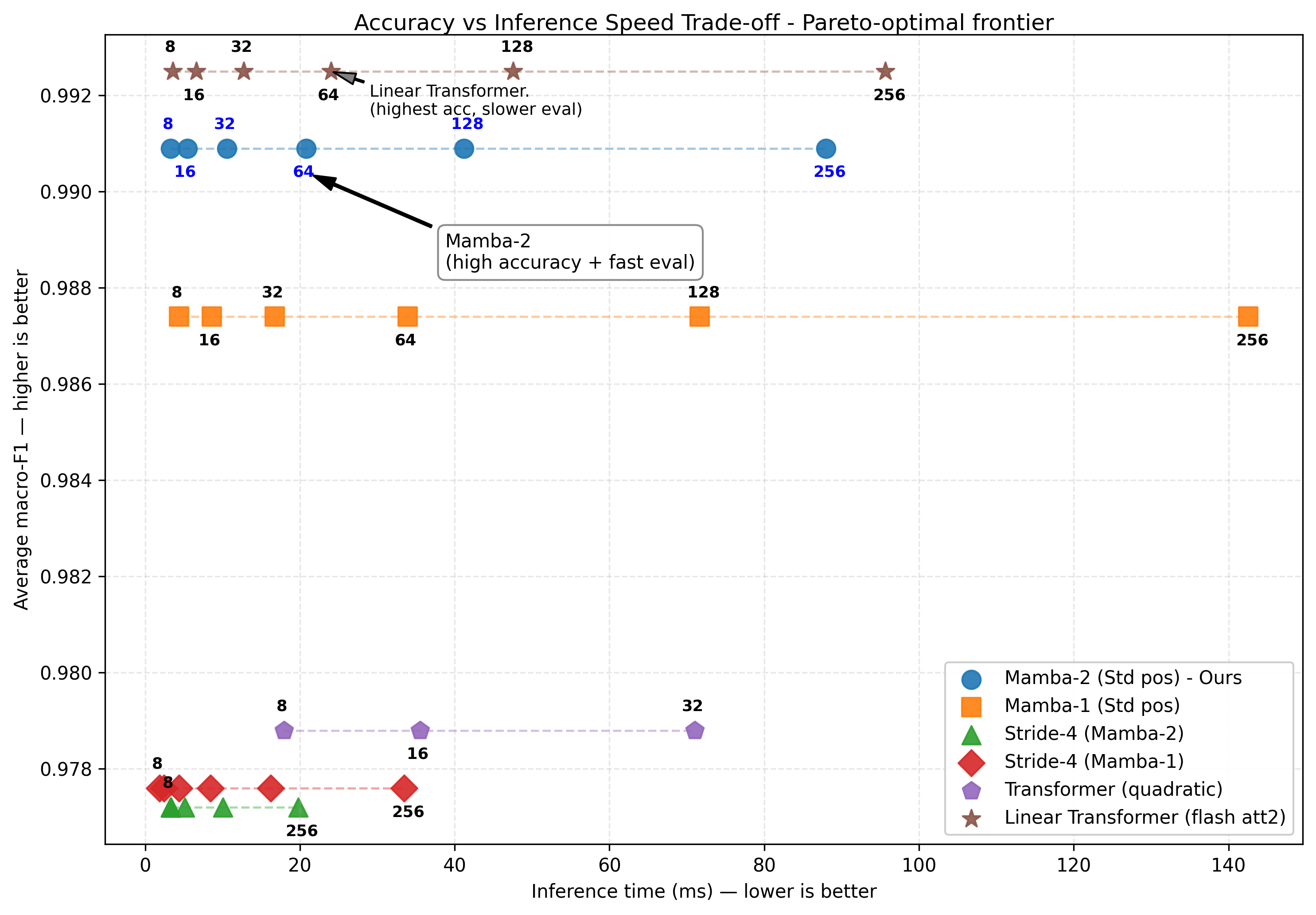}
    \caption{
    Comparison of average macro-F1 accuracy against evaluation time (ms) for different models across batch sizes 8, 16, 32, 64, 128, 256.
    \acrshort{OURMODEL} using Mamba-2 provides the best Pareto-optimal frontier, combining near-maximum accuracy with dramatically reduced inference latency compared to even Linear Transformers (with flash attention 2). Stride based model F1 scores are the lowest.
    All models use $d\_model=256, mlp=512, layers=4$.}
    \label{fig_acc_vs_performance}
\end{figure*}

Figure~\ref{fig_acc_vs_performance} shows macro-F1 vs inference time for batch sizes 8,16,32,64,128 and 256. Mamba-2 (Std pos) is Pareto optimal
\footnote{
$x^* \in S$ is Pareto optimal if and only if
$ \nexists\, x \in S : \Bigl( f_i(x) \leq f_i(x^*) \;\; \forall i \Bigr) \;\land\; \Bigl( \exists j : f_j(x) < f_j(x^*) \Bigr).$
}
, achieving near state-of-the-art accuracy (0.9909) while maintaining substantially faster inference times than the linear Transformer with FlashAttention-2 (highest F1 at 0.9925 but noticeably slower inference) and dramatically outperforms the vanilla Transformer and Mamba-1 in speed. While Stride-4 offers the fastest inference, its F1 is the lowest. 
Mamba-2 delivers substantially faster backward passes (often 30–60\% over Mamba-1 and 2–3× over the linear Transformer at medium-to-large batch sizes) and uses 2–4× less GPU memory, enabling larger effective batch sizes and faster ablation cycles on commodity hardware such as a single RTX 3090. (See Appendix Table~\ref{tab:mamba-scaling-d256-v2}.)
These results highlight Mamba-2's superior practical balance of predictive performance and inference efficiency for byte-level network burst classification even over highly optimized linear attention architectures when deployability and training throughput are prioritized.


\section{Discussion}

\subsection{Direct byte-level supervised learning is sufficient for strong traffic classification.}

A central finding of this work is that Mamba-based linear-time sequence modeling can learn discriminative traffic representations directly from raw packet bytes in a fully supervised setting. 
This contrasts with the dominant trend in recent traffic classification, where strong performance is often associated with heavy pre-training, large engineered representations, or both. 
As summarized in Table~\ref{tbl:rep-comp}, transformer based methods such as ET-BERT~\cite{lin2022etbert}, YaTC~\cite{zhao2023yatc}, and prior Mamba approaches such as NetMamba~\cite{wang2024netmamba}, NetMamba+~\cite{wang2026netmambaplus} rely on pre-training to obtain general-purpose traffic representations before downstream fine-tuning. 
Classical machine learning methods employing statistical features, such as AppScanner~\cite{taylor2017robust}, FlowPrint~\cite{van2020flowprint} also rely on pre-training. 
Furthermore, even supervised deep learning for traffic analysis using packet lengths or raw bytes such as FS-Net~\cite{liu2019fs}, TFE-GNN~\cite{zhang2023tfe} rely on pre-training and handcrafted features.
We show that by avoiding early patch/stride/token aggregation, preserving and providing undiluted fine-grained byte-resolution information to the Mamba backbone eliminates the need for an entire pretraining stage and can match or exceed much heavier pre-trained alternatives.

\subsection{Mamba-2's constrained transition structure is adequate for byte-level traffic modeling.}

Mamba-2 differs from Mamba-1 in that its state transition ($A$-matrix) is more restricted and constrained: the core transition matrix takes a scalar-times-identity form rather than a more flexible full diagonal matrix. In principle, this reduces the diversity of intrinsic time constants that can be represented directly within the transition dynamics.

By governing the state transition, i.e., how information is retained/decays/oscillates over time, the $A$-matrix controls how time (sequential) dynamics are represented in the SSM. In Mamba-1 (diagonal $A$), each hidden channel/state dimension can have its own decay rate/time constant, and the model can represent a mixture of many different memory scales simultaneously (fast, medium, slow) in parallel~\cite{gu2023mamba}.
Thus, in Mamba-1, each state dimension may learn its own decay behavior, naturally supporting a mixture of fast and slow timescales.
In contrast, in Mamba-2 $A$ is a 'scalar $\times$ I' or $A=\alpha I$, thus all state dimensions share essentially the same base decay rate (within a head/block). While more GPU-friendly and efficient, it is more restrictive. 
The diversity of dynamics must come from other factors (multi-head structure, learned projections, input-dependent parameters, gating, etc.)~\cite{dao2024Mamba-2}.
Thus, in Mamba-2, diversity must instead emerge through other components, including multi-head structure, learned projections, input-conditioned parameters, local convolution, and gating~\cite{dao2024Mamba-2}.
  
This design trade-off is particularly relevant for network traffic, which contains compositional 'multi-timescale' structures: (a) local-local byte patterns within packets, (b) local-to-medium packet boundaries, (c) medium range handshake/message sequences such as TLS and short protocol markers, and (d) mid-range handshake or message sequences, and (d) long-range flow-level behavior. 
One might therefore expect Mamba-1's more flexible diagonal $A$ to be a natural way to allocate different channels to different time constants.

However, our experiments show that this is not necessary in practice for burst-level byte classification. Under matched settings, Mamba-2 matches or exceeds Mamba-1 on nearly all evaluated datasets.
We show that for network byte modality the rest of Mamba-2 (multi-head structure + mixing + selective gating) compensates for the simpler base $A$ matrix.
This suggests that the remaining degrees of freedom in Mamba-2 are sufficient to capture the relevant multi-timescale behavior of packet-byte signals.

\subsection{The main empirical bottleneck is not transition flexibility, but early information loss.}
Our ablation study shows that the most damaging change is not the choice between Mamba-1 and Mamba-2, but the use of early aggregation and summarization from patches/strides or torkenization . 
The downsampling of the input sequence substantially reduces average performance and increases cross-dataset variance, indicating that fine-grained byte order carries critical discriminative information. 
This observation aligns with prior byte-level modeling work arguing that fixed patching or striding can obscure important structure~\cite{pagnoni2025byte}. In the network setting, such structure may include local protocol signatures, packet header organization, and short payload motifs. Preserving undiluted byte resolution appears to be more important than maximizing the flexibility of the latent dynamical system.

\subsection{Compact SSMs are a good fit for packet-byte classification.}
Our experiments show that strong performance does not require large latent state sizes or deep backbones. Moderate state sizes ($d_{\text{state}}=16$--$64$) and relatively shallow models already perform extremely well, while overly large states or aggressively reduced widths tend to hurt robustness. This indicates that the discriminative structure in packet bursts can be captured with compact selective SSMs, making the approach attractive for practical deployment scenarios where compute and memory are limited.

\subsection{Mamba-2 provides both modeling and systems advantages.}
Our scaling study shows that Mamba-2 delivers clear computational benefits over Mamba-1 in this application, especially during backpropagation. These efficiency gains matter because supervised training, benchmarking, and ablation studies all depend heavily on turnaround time. Taken together with the classification results, the evidence suggests that Mamba-2 is a particularly suitable backbone for byte-level burst classification: it preserves the linear-time advantages of SSMs, achieves strong predictive performance, and provides better training efficiency than Mamba-1 under comparable settings.

\subsection{Avoiding Self-Supervised Pre-training}
One of the key advantages of \acrshort{OURMODEL} is that it completely eliminates the need for self-supervised pre-training, a dominant but costly component in most state-of-the-art traffic classification pipelines.
Quantitatively, removing the pre-training stage yields major efficiency gains. In representative baselines such as ET-BERT, YaTC, and NetMamba, the pre-training phase typically consumes 10--100$  \times  $ more compute than the downstream fine-tuning stage, resulting in an estimated 3--15$  \times  $ reduction in total wall-clock training time on commodity GPUs.
Training memory footprint is simultaneously reduced by a factor of 2--4$  \times  $  relative to Transformer-based counterparts (see scaling tables), enabling larger effective batch sizes and single-GPU operation even at sequence length 1600.
We further incur zero risk of negative transfer from mismatched pre-training corpora, a well-known failure mode in traffic analysis under concept drift. 
From a non-quantifiable perspective, it collapses the experimental pipeline from two complex stages to a single end-to-end task, drastically lowers the hyperparameter-tuning burden (masking ratios, reconstruction objectives, or auxiliary losses), simplifies code maintenance and reproducibility, and eases real-world deployment in resource-constrained or rapidly evolving NIDS environments where frequent retraining is required. 
Overall, this positions direct byte-level supervised learning using compact selective state-space Mamba-2 as a markedly more practical and deployable alternative to the pre-training-heavy paradigm that has dominated recent literature.

\section{Conclusion}
We present \acrshort{OURMODEL}, a compact, tokenizer-free byte-level sequence classifier for network burst classification built on a Mamba-2 backbone. Unlike most recent strong baselines in encrypted traffic analysis and NIDS, our approach operates directly on raw packet bytes, avoids tokenization, patching, and heavy engineered multimodal inputs, and requires no self-supervised pre-training stage. Across six public benchmarks spanning encrypted mobile app identification, VPN/Tor classification, malware traffic classification, and IoT attack traffic, \acrshort{OURMODEL} achieves consistently strong performance and is competitive with, or superior to, substantially heavier and often pre-trained baselines. These results show that direct byte-to-classification learning is not only feasible for network traffic, but can be highly effective when paired with a suitable linear-time sequence model.


Our experiments show (1) preserving undiluted byte-level temporal resolution is critical for performance, early downsampling through striding causes severe degradation; (2) Mamba-2's more constrained transition dynamics not only suffice for this domain but can act as a beneficial regularizer, yielding improved consistency across datasets compared to Mamba-1; and (3) moderate model width and state sizes of 16-64, much lower that the defaults, are sufficient for robust generalization. Mamba-2 offers clear efficiency advantages, with 30--60\% faster backward passes and substantially lower memory usage than Mamba-1 or Transformer variants, enabling faster training and inference on commodity GPUs, without pre-training overhead.
These findings challenge the prevailing assumption that expensive pre-training pipelines are necessary for state-of-the-art traffic classification. 
Byte-level packet classification does not inherently require tokenization, large latent state sizes, or expensive pre-training pipelines, offering a simpler, more efficient, and more deployable paradigm for network traffic analysis. We believe this work opens promising new directions for byte-level SSMs in cybersecurity, including extensions to longer flows, online inference, concept-drift robustness, and edge-device deployment.

\subsection*{Acknowledgments}
Dedicated to Sugandi.

\onecolumn
\bibliographystyle{IEEEtran}
\bibliography{main}

\section*{Appendix}

\begin{table*}[h]
    \centering
    \small
    \caption{Scaling Comparison: Mamba-1 vs Mamba-2 at Larger Batch Sizes. $d\_model=256, mlp=512, layers=4$(default). Times in ms. Peak memory in MiB.
    Averaged over 10 runs, each with a warm-up of 10. Implementation: mamba\_ssm. OOM indicates out of memory on RTX 3090, 23.54GiB.
    Note As Flash attention in linear transformer uses $torch.autocast$ resulting in lower precision (fp16 and bf16), all models evaluated under torch AMP (automatic mixed precision).}
    \label{tab:mamba-scaling-d256-v2}
    \resizebox{\textwidth}{0.5\textheight}{
    \begin{tabular}{l r rrrrrrr}
    \toprule
    \multirow{2}{*}{\textbf{Architecture}} & \multirow{2}{*}{\textbf{Param(M)}} & \multirow{2}{*}{\textbf{Batch}} & \multicolumn{3}{c}{\textbf{Time (ms)}} & \multicolumn{3}{c}{\textbf{Peak Memory (MiB)}} \\ 
    \cmidrule(lr){4-6} \cmidrule(lr){7-9} 
    & & & {Forward} & {Back} & {Eval} & {Forward} & {Back} & {Eval} \\ 
    \midrule
    Std pos & 2.5 & 8 & 3.69 & 9.71 & 3.28 & 364.02 & 476.56 & 164.98 \\ 
Std pos (Mamba-1) & 2.6 & 8 & 4.35 & 13.28 & 4.38 & 395.61 & 455.60 & 190.68 \\ 
Stride (4) & 2.2 & 8 & 3.51 & 9.34 & 3.26 & 144.80 & 185.01 & 110.62 \\ 
Stride (4,Mamba-1) & 2.3 & 8 & 2.20 & 4.54 & 1.86 & 165.74 & 183.07 & 129.55 \\ 
No emb proj & 2.2 & 8 & 3.57 & 9.56 & 3.23 & 398.01 & 511.18 & 240.53 \\ 
4 layers & 2.5 & 8 & 3.69 & 10.11 & 3.30 & 461.86 & 574.39 & 261.13 \\ 
1 layers & 1.3 & 8 & 1.42 & 3.82 & 1.09 & 301.93 & 414.46 & 256.87 \\ 
d\_state 32 & 2.5 & 8 & 3.79 & 10.15 & 3.32 & 496.84 & 613.76 & 294.22 \\ 
d\_state 64 & 2.6 & 8 & 3.71 & 10.13 & 3.33 & 527.68 & 652.96 & 321.67 \\ 
d\_state 128 & 2.7 & 8 & 3.75 & 10.32 & 3.39 & 571.56 & 712.35 & 355.45 \\ 
Transformer & 2.5 & 8 & 21.67 & 26.85 & 17.97 & 3194.19 & 3332.52 & 783.94 \\ 
Linear transformer & 2.5 & 8 & 3.67 & 8.93 & 3.56 & 714.78 & 714.78 & 339.63 \\ 
\midrule 
Std pos & 2.5 & 16 & 5.32 & 13.68 & 5.45 & 911.62 & 1108.89 & 497.67 \\ 
Std pos (Mamba-1) & 2.6 & 16 & 8.52 & 30.62 & 8.62 & 936.27 & 1055.16 & 506.89 \\ 
Stride (4) & 2.2 & 16 & 3.65 & 9.52 & 3.27 & 383.69 & 447.70 & 309.48 \\ 
Stride (4,Mamba-1) & 2.3 & 16 & 2.64 & 5.48 & 2.44 & 392.05 & 425.63 & 312.30 \\ 
No emb proj & 2.2 & 16 & 4.77 & 12.52 & 4.89 & 823.79 & 1020.06 & 496.56 \\ 
4 layers & 2.5 & 16 & 5.32 & 13.65 & 5.45 & 912.11 & 1109.38 & 497.39 \\ 
1 layers & 1.3 & 16 & 2.00 & 5.00 & 1.94 & 566.94 & 764.21 & 471.33 \\ 
d\_state 32 & 2.5 & 16 & 5.40 & 13.86 & 5.51 & 922.80 & 1127.14 & 502.41 \\ 
d\_state 64 & 2.6 & 16 & 5.82 & 14.76 & 5.94 & 940.79 & 1161.57 & 513.52 \\ 
d\_state 128 & 2.7 & 16 & 6.49 & 16.73 & 6.61 & 979.90 & 1232.80 & 535.08 \\ 
Transformer & 2.5 & 16 & 43.26 & 51.70 & 35.54 & 6185.40 & 6461.41 & 1354.12 \\ 
Linear transformer & 2.5 & 16 & 6.61 & 15.78 & 6.59 & 1190.87 & 1190.87 & 427.73 \\ 
\midrule 
Std pos & 2.5 & 32 & 10.36 & 25.35 & 10.57 & 1571.90 & 1964.18 & 742.45 \\ 
Std pos (Mamba-1) & 2.6 & 32 & 16.13 & 74.69 & 16.76 & 1611.04 & 1847.63 & 761.45 \\ 
Stride (4) & 2.2 & 32 & 3.69 & 9.71 & 3.33 & 527.36 & 654.59 & 371.14 \\ 
Stride (4,Mamba-1) & 2.3 & 32 & 4.44 & 10.12 & 4.37 & 544.86 & 610.96 & 375.98 \\ 
No emb proj & 2.2 & 32 & 9.38 & 23.30 & 9.57 & 1397.49 & 1790.39 & 744.55 \\ 
4 layers & 2.5 & 32 & 10.36 & 25.36 & 10.57 & 1574.63 & 1967.98 & 747.77 \\ 
1 layers & 1.3 & 32 & 3.77 & 8.89 & 3.80 & 891.71 & 1285.36 & 696.18 \\ 
d\_state 32 & 2.5 & 32 & 10.49 & 25.90 & 10.69 & 1602.91 & 2011.12 & 758.89 \\ 
d\_state 64 & 2.6 & 32 & 11.35 & 27.67 & 11.49 & 1639.76 & 2080.56 & 780.65 \\ 
d\_state 128 & 2.7 & 32 & 12.73 & 31.73 & 12.90 & 1719.38 & 2224.92 & 823.52 \\ 
Transformer & 2.5 & 32 & 86.41 & 102.11 & 71.03 & 12129.91 & 12681.76 & 2460.00 \\ 
Linear transformer & 2.5 & 32 & 13.00 & 30.16 & 12.76 & 2161.72 & 2161.72 & 607.01 \\ 
\midrule 
Std pos & 2.5 & 64 & 20.80 & 48.12 & 20.80 & 2907.68 & 3691.31 & 1237.62 \\ 
Std pos (Mamba-1) & 2.6 & 64 & 31.26 & 156.31 & 33.94 & 2981.16 & 3452.22 & 1276.67 \\ 
Stride (4) & 2.2 & 64 & 5.00 & 13.15 & 5.10 & 816.58 & 1039.42 & 489.64 \\ 
Stride (4,Mamba-1) & 2.3 & 64 & 8.37 & 19.74 & 8.42 & 846.53 & 978.50 & 500.13 \\ 
No emb proj & 2.2 & 64 & 18.48 & 43.97 & 18.82 & 2553.30 & 3336.44 & 1236.38 \\ 
4 layers & 2.5 & 64 & 20.77 & 47.79 & 20.88 & 2903.16 & 3686.94 & 1235.64 \\ 
1 layers & 1.3 & 64 & 7.39 & 16.38 & 7.55 & 1531.21 & 2314.84 & 1133.96 \\ 
d\_state 32 & 2.5 & 64 & 21.09 & 48.92 & 21.01 & 2945.52 & 3760.53 & 1259.53 \\ 
d\_state 64 & 2.6 & 64 & 22.34 & 53.79 & 22.68 & 3027.67 & 3907.00 & 1302.30 \\ 
d\_state 128 & 2.7 & 64 & 25.04 & 60.62 & 25.48 & 3189.62 & 4196.85 & 1389.24 \\ 
Transformer & 2.5 & 64 & OOM & OOM & OOM & OOM & OOM & OOM \\ 
Linear transformer & 2.5 & 64 & 24.33 & 56.68 & 24.02 & 4018.61 & 4018.61 & 957.22 \\ 
\midrule 
Std pos & 2.5 & 128 & 40.70 & 95.35 & 41.17 & 5568.93 & 7134.06 & 2219.62 \\ 
Std pos (Mamba-1) & 2.6 & 128 & 68.26 & 313.53 & 71.66 & 5716.04 & 6655.32 & 2297.93 \\ 
Stride (4) & 2.2 & 128 & 9.74 & 24.33 & 10.04 & 1383.47 & 1826.24 & 727.33 \\ 
Stride (4,Mamba-1) & 2.3 & 128 & 16.03 & 40.63 & 16.23 & 1444.04 & 1705.94 & 748.99 \\ 
No emb proj & 2.2 & 128 & 36.58 & 86.11 & 37.33 & 4865.29 & 6430.75 & 2219.34 \\ 
4 layers & 2.5 & 128 & 40.56 & 95.22 & 41.20 & 5566.41 & 7131.54 & 2219.71 \\ 
1 layers & 1.3 & 128 & 14.54 & 32.46 & 14.89 & 2815.73 & 4380.90 & 2018.03 \\ 
d\_state 32 & 2.5 & 128 & 41.19 & 96.98 & 41.78 & 5647.66 & 7277.54 & 2260.63 \\ 
d\_state 64 & 2.6 & 128 & 44.46 & 106.05 & 45.07 & 5810.08 & 7568.36 & 2347.53 \\ 
d\_state 128 & 2.7 & 128 & 49.96 & 120.06 & 50.68 & 6131.53 & 8145.29 & 2520.49 \\ 
Transformer & 2.5 & 128 & OOM & OOM & OOM & OOM & OOM & OOM \\ 
Linear transformer & 2.5 & 128 & 48.56 & 107.18 & 47.53 & 7778.24 & 7778.24 & 1657.55 \\ 
\midrule 
Std pos & 2.5 & 256 & 81.71 & 184.59 & 87.96 & 10892.18 & 14022.54 & 4182.62 \\ 
Std pos (Mamba-1) & 2.6 & 256 & 142.56 & 624.64 & 142.56 & 11185.15 & 13062.64 & 4338.48 \\ 
Stride (4) & 2.2 & 256 & 19.45 & 46.30 & 19.77 & 2520.27 & 3404.49 & 1198.21 \\ 
Stride (4,Mamba-1) & 2.3 & 256 & 31.34 & 89.58 & 33.49 & 2635.81 & 3157.05 & 1241.11 \\ 
No emb proj & 2.2 & 256 & 74.08 & 169.53 & 75.54 & 9487.40 & 12617.04 & 4181.06 \\ 
4 layers & 2.5 & 256 & 81.82 & 184.24 & 87.84 & 10889.76 & 14019.57 & 4179.79 \\ 
1 layers & 1.3 & 256 & 28.92 & 61.77 & 29.70 & 5379.59 & 8509.31 & 3777.90 \\ 
d\_state 32 & 2.5 & 256 & 82.97 & 188.53 & 89.14 & 11049.19 & 14306.66 & 4264.63 \\ 
d\_state 64 & 2.6 & 256 & 89.51 & 209.09 & 92.47 & 11370.64 & 14884.23 & 4436.43 \\ 
d\_state 128 & 2.7 & 256 & 100.38 & 239.32 & 101.51 & 12013.63 & 16039.47 & 4781.30 \\ 
Transformer & 2.5 & 256 & OOM & OOM & OOM & OOM & OOM & OOM \\ 
Linear transformer & 2.5 & 256 & 97.73 & 208.76 & 95.67 & 15311.50 & 15311.50 & 3056.88 \\ 
\midrule 
\bottomrule
    \end{tabular}
    }
    \end{table*}

\begin{table*}[h]
    \centering
    \small
    \caption{Scaling Comparison: Mamba-1 vs Mamba-2 at Larger Batch Sizes. $d\_model=64, mlp=64, layers=2$(default). Times in ms. Peak memory in MiB.
    Averaged over 10 runs, each with a warm-up of 10. Implementation: mamba\_ssm. OOM indicates out of memory on RTX 3090, 23.54GiB.
    Note As Flash attention in linear transformer uses $torch.autocast$ resulting in lower precision (fp16 and bf16), all models evaluated under torch AMP (automatic mixed precision).}
    \label{tab:mamba-scaling-d64-v2}
    \resizebox{\textwidth}{0.5\textheight}{
    \begin{tabular}{l r rrrrrrr}
    \toprule
    \multirow{2}{*}{\textbf{Architecture}} & \multirow{2}{*}{\textbf{Param(M)}} & \multirow{2}{*}{\textbf{Batch}} & \multicolumn{3}{c}{\textbf{Time (ms)}} & \multicolumn{3}{c}{\textbf{Peak Memory (MiB)}} \\ 
    \cmidrule(lr){4-6} \cmidrule(lr){7-9} 
    & & & {Forward} & {Back} & {Eval} & {Forward} & {Back} & {Eval} \\ 
    \midrule
    Std pos & 0.2 & 8 & 2.25 & 5.24 & 1.86 & 105.51 & 154.54 & 62.29 \\ 
Std pos (Mamba-1) & 0.3 & 8 & 1.40 & 2.66 & 1.11 & 99.59 & 116.05 & 56.25 \\ 
Stride (4) & 0.2 & 8 & 2.09 & 5.06 & 1.80 & 35.81 & 54.43 & 33.01 \\ 
Stride (4,Mamba-1) & 0.2 & 8 & 1.34 & 2.52 & 1.06 & 35.00 & 39.55 & 31.88 \\ 
No emb proj & 0.2 & 8 & 2.01 & 5.09 & 1.76 & 81.00 & 130.07 & 68.10 \\ 
4 layers & 0.3 & 8 & 3.76 & 9.66 & 3.38 & 145.37 & 194.44 & 70.62 \\ 
1 layers & 0.2 & 8 & 1.43 & 3.29 & 1.10 & 99.71 & 148.78 & 68.28 \\ 
d\_state 32 & 0.3 & 8 & 2.13 & 5.56 & 1.88 & 119.25 & 171.66 & 75.12 \\ 
d\_state 64 & 0.3 & 8 & 2.22 & 5.48 & 1.91 & 126.47 & 185.58 & 80.81 \\ 
d\_state 128 & 0.3 & 8 & 2.16 & 5.39 & 1.90 & 207.24 & 279.82 & 147.37 \\ 
Transformer & 0.3 & 8 & 10.26 & 12.14 & 8.27 & 1608.67 & 1759.58 & 528.26 \\ 
Linear transformer & 0.3 & 8 & 1.29 & 2.83 & 0.97 & 154.87 & 161.31 & 71.46 \\ 
\midrule 
Std pos & 0.2 & 16 & 2.23 & 5.78 & 1.90 & 208.86 & 306.95 & 122.22 \\ 
Std pos (Mamba-1) & 0.3 & 16 & 1.60 & 4.89 & 1.41 & 195.29 & 228.17 & 106.10 \\ 
Stride (4) & 0.2 & 16 & 2.11 & 5.20 & 1.90 & 67.75 & 103.16 & 61.13 \\ 
Stride (4,Mamba-1) & 0.2 & 16 & 1.36 & 2.47 & 1.09 & 63.52 & 73.48 & 56.65 \\ 
No emb proj & 0.2 & 16 & 2.07 & 5.18 & 1.80 & 149.42 & 247.51 & 122.10 \\ 
4 layers & 0.3 & 16 & 3.84 & 10.12 & 3.47 & 268.95 & 367.04 & 122.30 \\ 
1 layers & 0.2 & 16 & 1.45 & 3.63 & 1.11 & 178.82 & 276.91 & 115.94 \\ 
d\_state 32 & 0.3 & 16 & 2.27 & 5.84 & 1.88 & 214.10 & 318.89 & 125.92 \\ 
d\_state 64 & 0.3 & 16 & 2.24 & 5.79 & 1.88 & 225.34 & 343.51 & 133.32 \\ 
d\_state 128 & 0.3 & 16 & 2.25 & 5.90 & 1.88 & 247.67 & 392.60 & 148.57 \\ 
Transformer & 0.3 & 16 & 20.18 & 22.96 & 16.41 & 3181.91 & 3485.72 & 1020.46 \\ 
Linear transformer & 0.3 & 16 & 1.87 & 4.51 & 1.64 & 268.68 & 281.36 & 99.50 \\ 
\midrule 
Std pos & 0.2 & 32 & 2.35 & 6.85 & 2.23 & 377.21 & 573.34 & 204.67 \\ 
Std pos (Mamba-1) & 0.3 & 32 & 2.79 & 10.99 & 2.75 & 350.92 & 416.64 & 172.41 \\ 
Stride (4) & 0.2 & 32 & 2.11 & 5.19 & 1.85 & 96.66 & 151.42 & 82.68 \\ 
Stride (4,Mamba-1) & 0.2 & 32 & 1.33 & 2.56 & 1.08 & 88.08 & 106.14 & 71.99 \\ 
No emb proj & 0.2 & 32 & 2.08 & 5.71 & 1.83 & 258.19 & 454.33 & 204.55 \\ 
4 layers & 0.3 & 32 & 3.89 & 11.15 & 3.74 & 497.49 & 693.70 & 204.74 \\ 
1 layers & 0.2 & 32 & 1.58 & 4.56 & 1.53 & 317.07 & 513.20 & 192.13 \\ 
d\_state 32 & 0.3 & 32 & 2.35 & 6.85 & 2.29 & 388.45 & 598.34 & 212.06 \\ 
d\_state 64 & 0.3 & 32 & 2.60 & 7.35 & 2.47 & 409.39 & 645.77 & 226.83 \\ 
d\_state 128 & 0.3 & 32 & 2.93 & 8.31 & 2.86 & 453.06 & 744.03 & 256.38 \\ 
Transformer & 0.3 & 32 & 38.54 & 42.91 & 31.00 & 6322.62 & 6930.17 & 1999.58 \\ 
Linear transformer & 0.3 & 32 & 3.51 & 8.04 & 3.15 & 502.76 & 502.76 & 159.11 \\ 
\midrule 
Std pos & 0.2 & 64 & 4.42 & 10.05 & 4.41 & 716.18 & 1109.25 & 369.57 \\ 
Std pos (Mamba-1) & 0.3 & 64 & 5.45 & 21.29 & 5.66 & 668.55 & 800.22 & 306.08 \\ 
Stride (4) & 0.2 & 64 & 2.11 & 5.31 & 1.87 & 151.76 & 261.22 & 124.97 \\ 
Stride (4,Mamba-1) & 0.2 & 64 & 1.51 & 3.30 & 1.34 & 135.19 & 171.27 & 105.12 \\ 
No emb proj & 0.2 & 64 & 3.37 & 8.39 & 3.30 & 477.56 & 869.82 & 369.44 \\ 
4 layers & 0.3 & 64 & 7.22 & 17.46 & 7.25 & 956.58 & 1350.31 & 369.64 \\ 
1 layers & 0.2 & 64 & 3.02 & 6.34 & 2.96 & 597.43 & 990.36 & 344.52 \\ 
d\_state 32 & 0.3 & 64 & 4.52 & 10.59 & 4.53 & 739.73 & 1158.74 & 384.33 \\ 
d\_state 64 & 0.3 & 64 & 4.89 & 11.47 & 4.91 & 783.17 & 1254.78 & 413.87 \\ 
d\_state 128 & 0.3 & 64 & 5.67 & 13.82 & 5.65 & 866.12 & 1446.16 & 473.90 \\ 
Transformer & 0.3 & 64 & 79.26 & 84.11 & 63.95 & 12608.90 & 13821.93 & 3959.27 \\ 
Linear transformer & 0.3 & 64 & 6.72 & 12.67 & 6.15 & 960.58 & 960.58 & 278.31 \\ 
\midrule 
Std pos & 0.2 & 128 & 8.62 & 18.79 & 8.71 & 1392.42 & 2176.99 & 699.36 \\ 
Std pos (Mamba-1) & 0.3 & 128 & 11.03 & 42.39 & 11.51 & 1278.08 & 1541.18 & 570.62 \\ 
Stride (4) & 0.2 & 128 & 2.22 & 5.78 & 1.91 & 262.84 & 481.72 & 208.65 \\ 
Stride (4,Mamba-1) & 0.2 & 128 & 2.42 & 6.03 & 2.34 & 229.61 & 301.72 & 169.76 \\ 
No emb proj & 0.2 & 128 & 6.62 & 15.74 & 6.64 & 916.99 & 1701.42 & 699.23 \\ 
4 layers & 0.3 & 128 & 14.25 & 33.14 & 14.39 & 1873.92 & 2658.39 & 699.43 \\ 
1 layers & 0.2 & 128 & 5.91 & 11.66 & 5.93 & 1151.70 & 1936.14 & 649.29 \\ 
d\_state 32 & 0.3 & 128 & 8.85 & 19.72 & 8.89 & 1436.49 & 2275.90 & 728.88 \\ 
d\_state 64 & 0.3 & 128 & 9.62 & 21.63 & 9.61 & 1518.72 & 2464.19 & 788.15 \\ 
d\_state 128 & 0.3 & 128 & 11.13 & 26.32 & 11.11 & 1687.40 & 2847.04 & 906.02 \\ 
Transformer & 0.3 & 128 & OOM & OOM & OOM & OOM & OOM & OOM \\ 
Linear transformer & 0.3 & 128 & 13.19 & 24.05 & 12.01 & 1880.54 & 1880.54 & 516.73 \\ 
\midrule 
Std pos & 0.2 & 256 & 16.92 & 36.63 & 17.02 & 2747.06 & 4318.06 & 1359.03 \\ 
Std pos (Mamba-1) & 0.3 & 256 & 21.32 & 84.31 & 22.30 & 2518.75 & 3044.30 & 1100.66 \\ 
Stride (4) & 0.2 & 256 & 3.50 & 8.72 & 3.44 & 487.27 & 925.69 & 377.65 \\ 
Stride (4,Mamba-1) & 0.2 & 256 & 4.70 & 11.50 & 4.75 & 425.00 & 568.34 & 298.22 \\ 
No emb proj & 0.2 & 256 & 13.05 & 30.21 & 13.05 & 1794.09 & 3366.64 & 1358.90 \\ 
4 layers & 0.3 & 256 & 27.94 & 64.61 & 28.00 & 3710.08 & 5280.28 & 1359.10 \\ 
1 layers & 0.2 & 256 & 11.34 & 22.58 & 11.40 & 2265.16 & 3834.67 & 1259.13 \\ 
d\_state 32 & 0.3 & 256 & 17.19 & 38.31 & 17.23 & 2832.97 & 4509.78 & 1418.07 \\ 
d\_state 64 & 0.3 & 256 & 18.59 & 42.36 & 18.84 & 3000.35 & 4890.60 & 1536.15 \\ 
d\_state 128 & 0.3 & 256 & 21.77 & 52.10 & 21.60 & 3336.10 & 5654.90 & 1772.49 \\ 
Transformer & 0.3 & 256 & OOM & OOM & OOM & OOM & OOM & OOM \\ 
Linear transformer & 0.3 & 256 & 25.41 & 46.96 & 23.09 & 3701.21 & 3701.21 & 993.56 \\ 
\midrule 
\bottomrule
    \end{tabular}
    }
    \end{table*}

\end{document}